\documentclass[%
 notitlepage,
 reprint,
 superscriptaddress,
 amsmath,amssymb,
 aps,
]{revtex4-2}

\usepackage{lipsum}
\usepackage{graphicx}
\usepackage{dcolumn}
\usepackage{bm}

\usepackage{epstopdf}
\usepackage{xcolor}

\newcommand{\bra}[1]{\langle #1 \vert}
\newcommand{\ket}[1]{\vert #1 \rangle}

\begin{document}

\preprint{APS/123-QED}

\title{Topological speckles} 

\author{Y. M. I. A. Rodrigues}
\affiliation{%
 Instituto de F\'{i}sica, Universidade Federal de Alagoas, 57072-900 Macei\'{o}, AL, Brazil
}%

\author{M. F. V. Oliveira}
\affiliation{%
 Instituto de F\'{i}sica, Universidade Federal de Alagoas, 57072-900 Macei\'{o}, AL, Brazil
}%

\author{A. M. C. Souza}
\affiliation{%
 Departamento de F\'{i}sica, Universidade Federal de Sergipe, 49100-000 S\~{a}o Crist\'{o}v\~{a}o, SE, Brazil
}%

\author{M. L. Lyra}
\affiliation{%
 Instituto de F\'{i}sica, Universidade Federal de Alagoas, 57072-900 Macei\'{o}, AL, Brazil
}%

\author{F. A. B. F. de Moura}
\affiliation{%
 Instituto de F\'{i}sica, Universidade Federal de Alagoas, 57072-900 Macei\'{o}, AL, Brazil
}%

\author{G. M. A. Almeida}
\email{gmaalmeida@fis.ufal.br}
\affiliation{%
 Instituto de F\'{i}sica, Universidade Federal de Alagoas, 57072-900 Macei\'{o}, AL, Brazil
}%

\begin{abstract}
The time evolution of a topological Su-Schrieffer-Heeger chain
is analyzed through the statistics of speckle patterns. The emergence of topological edge states
dramatically affects the dynamical fluctuations of the wavefunction.
The intensity statistics is found to be described by a family of noncentral chi-squared distributions,
with the noncentrality parameter reflecting on the degree of edge-state localization. The response
of the speckle contrast with respect to the dimerization of the chain is explored in detail as well as 
the role of chiral symmetry-breaking disorder, number of edge states, their energy gap, and the locations between which the transport occurs. 
In addition to providing a venue for speckle customization, our results appeal to the use of speckle patterns
for characterization of nontrivial topological properties.
\end{abstract}

\maketitle

\textit{Introduction.}
Speckles are the result of interference between many wave components \cite{goodmanbook}, manifested as those
well-known random granular textures in images for example. 
But what it may seem like meaningless noise can often 
be associated to a well defined universal fluctuation pattern. Indeed, speckles meet a wide range of  
applications which includes medical imaging \cite{heeman19}, 
biosensing \cite{verwohlt18},
surface characterization \cite{zhang23},
optical metrology \cite{luo21}, 
and manipulation of cold atoms \cite{jendrzejewski12, delande14}, 
to name a few.
Hence, there has been a considerable effort directed toward the development of techniques for speckle customization  \cite{bromberg14,han23, bender23}.

The study of speckle patterns 
also carries a more fundamental appeal. It
can disclose valuable information about the underlying system that otherwise would be very difficult to probe \cite{verwohlt18,zhang23}. For instance, it has been shown recently that the entanglement due to symmetrization of identical bosons and fermions in a lattice is imprinted in the speckle behavior of their time-evolving wavefunction in the form of non-Rayleigh statistics \cite{oliveira23}. 
Speckle theory can thus
be used to explore 
involved quantum phenomena, such as entanglement \cite{beenakker09,peeters10,pires12,oliveira23}, many-body quantum dynamics \cite{kirkby22}, and long-tailed extreme events \cite{buarque22,buarque22-2}. These latter studies have shed new light on the role of disorder in producing events like rogue waves in low-dimensional quantum chains.    

In this letter we 
turn our attention to the prospect of obtaining fingerprints of topological edge states by analyzing 
speckle patterns. Topological lattices have been subject of intense research over the years (see \cite{ozawa19} and refs. therein).
Their robustness against disorder and rich transport properties associated to nontrivial topological phases of matter
meet the convenience photonic implementations \cite{kang23}. 

Here we consider a topological chain of the Su-Schrieffer-Heeger (SSH) type \cite{su79}. It follows a dimerized pattern of weak and strong couplings.
A single or a pair of edges states emerge due to the bulk-boundary correspondence depending on the parity of the number of sites.
By mapping the unitary Hamiltonian evolution onto random phasor models we derive a family of noncentral chi-squared distributions that describes local intensity fluctuations in the SSH model.   
The localization degree of the edge states ultimately controls
the speckle contrast. The influence of symmetry-breaking disorder and input-output locations are also analyzed. 


\begin{figure}[t!] 
\includegraphics[width=0.4\textwidth]{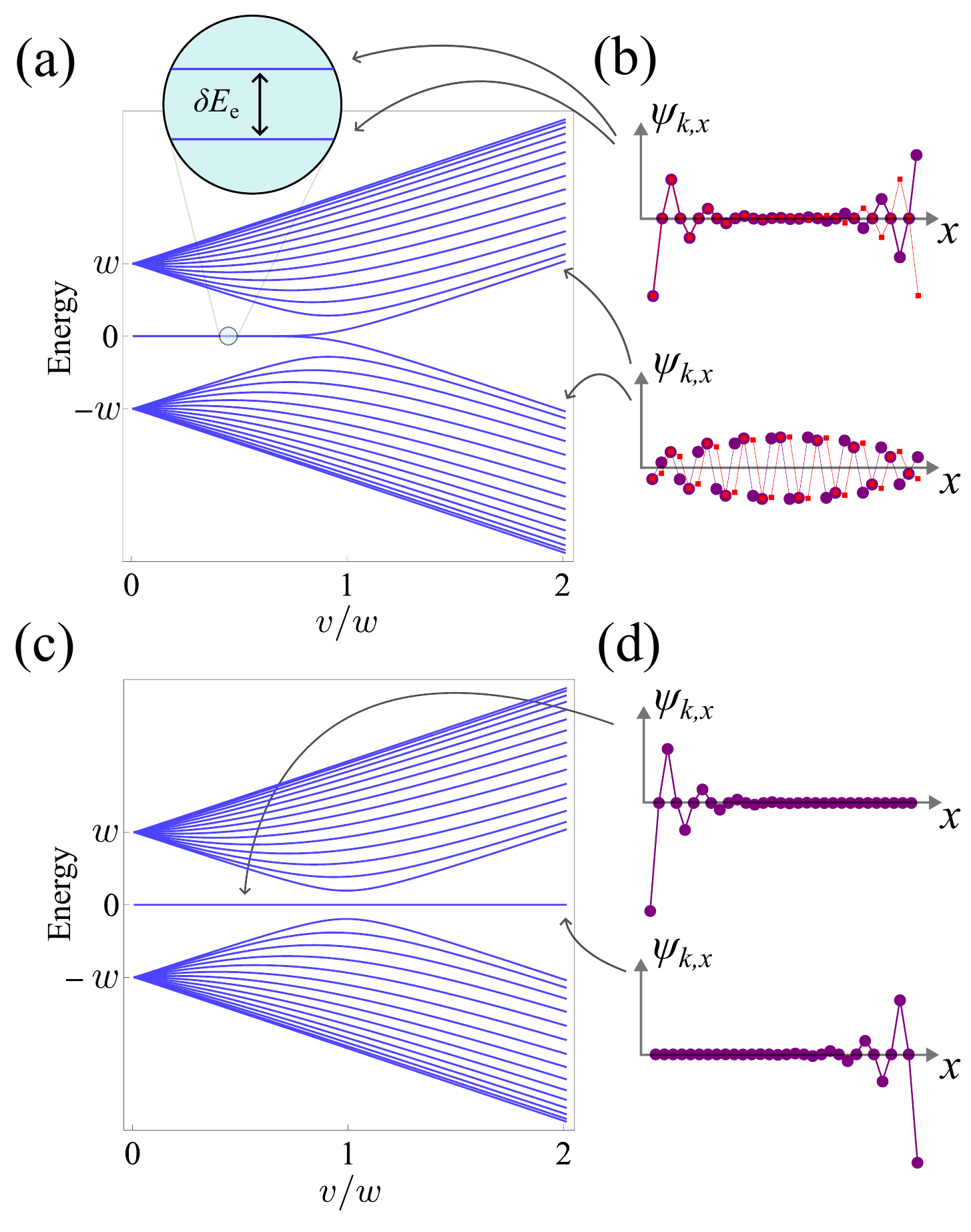}
\caption{\label{fig1} 
(a) Energy levels of the SSH chain 
versus $v/w$ for $N=30$. The inset highlights
the small gap $\delta E_{\mathrm{e}}=E_{\mathrm{e+}}-E_{\mathrm{e-}}$ 
between the edge-state modes. (b) Upper panel displays the 
wavefunction spatial profile of the
edge states
for $v/w =1/2$ whereas the lower panel depicts the same pair of modes when
$v/w=2$, to which the localization properties no longer holds. 
(c) Energy spectrum for $N=31$. When $N$ is odd, there is a single 
edge state at the middle of the band
strongly localized in either ends of the chain, as shown in (d) for $v/w =1/2$ 
and $v/w =2$ (upper and lower panels, respectively). 
}
\end{figure}

\textit{Model.} Consider a 1D staggered $N-$site chain with open boundaries described
by the tight-binding model ($\hbar=1$)
\begin{equation}
H=\sum_{x=1}^{N}\epsilon_{x}\ket{x}\bra{x}
+\sum_{x=1}^{N-1}J_{x,x+1}(\ket{x+1}\bra{x}+\mathrm{H.c.}),
\end{equation}
where $\ket{x}$ denotes a single particle occupying the $x$-th site, $\epsilon_x$ is the associated on-site energy, and $J_{x,x+1}$ is the hopping strength between neighboring sites. 
We now set a dimerized pattern of hopping rates such that $J_{x,x+1}=v$ for odd $x$ and $J_{x,x+1}=w$ for even $x$ (we use units such that $w=1$).  

For now, let us assume a homogeneous chain by setting $\epsilon_x=\epsilon=0$, for simplicity. The resulting Hamiltonian is the standard SSH model. In the thermodynamic limit (or assuming periodic conditions)  the model 
supports two insulating (gapped) phases when $v<w$ or $v>w$ that are topologically equivalent as characterized by distinct winding numbers (a clear and concise review on the matter can be found in \cite{batra20}). The hallmark of the model is the topological phase transition that occurs by crossing the metallic (gapless) phase at $v=w$.

In an open chain the bulk-boundary correspondence entails the existence of edge states. Indeed, for even $N$ one of those distinct topological phases (the nontrivial one; $v<w$) corresponds to a pair of nearly degenerate
eigenstates in the middle of the band gap strongly localized at both edges of the chain [see Figs. \ref{fig1}(a) and \ref{fig1}(b)].
The gap between their energies $\delta E_{\mathrm{e}}=E_{\mathrm{e+}}-E_{\mathrm{e-}}$ closes rapidly 
as $\sim (v/w)^{N/2+1}$ \cite{almeida18}.  
When $N$ is odd, although no topological transition occurs, a single zero-energy edge state emerges, now strongly localized at one of the ends of the chain according to $v<w$ or $v>w$ [see Figs. \ref{fig1}(c) and \ref{fig1}(d)]. Our goal here is to harness those edge states with the purpose of generating speckles with tailored contrasts. 
Conversely, the behavior of the contrast in response to the dimerization of the chain unveils
a subtle interplay between the edge modes, the existence of a gap between them, and on-site disorder. 



\textit{Edge-state phasor.}
Let us begin our analysis by focusing on the odd $N$ case. From now on we consider $v<w$.
The corresponding zero-energy edge state $\psi_{\mathrm{e},x}$ can be found analytically \cite{ciccarello11} and reads
\begin{equation} \label{bound}
\psi_{\mathrm{e},2m-1}= (-1)^{m}v^{m-1}\sqrt{\frac{v^2-1}{v^{N+1}-1}},
\end{equation}
for $m=1,\ldots,(N+1)/2$ and $v\in(0,1)$.
Note that it only has components at odd sites.

The speckle generation is provided 
by the unitary, Hamiltonian-based evolution of the system. As such, given a delta-like input $\ket{\Psi(t=0)}=\ket{x_{0}}$, the wavefunction
at site $x$ at a later time $t$ is
\begin{equation}\label{phasor}
\Psi_x(t) = \psi_{\mathrm{e},x_0}\psi_{\mathrm{e},x}+\sum_{k\neq\mathrm{e}}\psi_{k,x_0}\psi_{k,x}e^{-iE_kt},
\end{equation}
where $E_{k}$ are the $M=N-1$ eigenvalues belonging to the two continuous bands [see Fig. \ref{fig1}(c)]. The second term of Eq. (\ref{phasor}) can be regarded as a random phasor sum for time steps $\Delta t >> 1/w$
as if the time-dependent phases $\phi_k(t)=E_{k}t$ (mod $2\pi$)
were random phases uniformly 
distributed in $(0,2\pi)$ \cite{oliveira23}. The amplitudes $C_k = C_k (x_0,x) \equiv \psi_{k,x_0}\psi_{k,x}$ ($k\neq \mathrm{e}$) are made from $M$ delocalized modes $\psi_{k,x}\sim 1/\sqrt{M}$ and thus $C_k\sim 1/M$.
The edge-state phasor $C_{\mathrm{e}}$ is calculated from Eq. (\ref{bound}). All amplitudes are assumed to be real \textcolor{black}{without loss of generality}. 

\begin{figure}[t!] 
\includegraphics[width=0.47\textwidth]{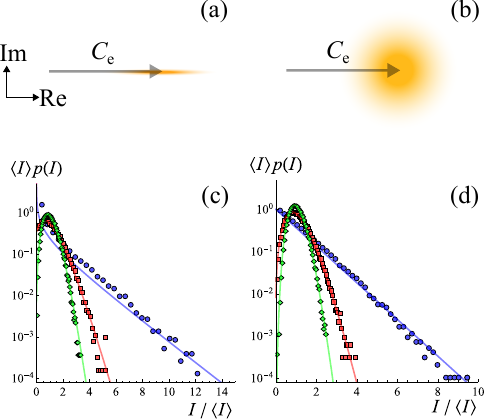}
\caption{\label{fig2} 
The phasor sum due to the truncated SSH Hamiltonian time evolution with odd $N$ can be seen as 
the constant edge-state phasor $C_{\mathrm{e}}$ perturbed by a (a) Gaussian noise on the real axis in the absence of the disorder or (b) by a circular Gaussian noise in the presence of disorder. 
This transition to a circular random variable occurs due to the
chiral symmetry breaking. Panels (c) and (d) show speckle intensity distributions for the clean and disordered (single realization) cases, respectively, considering $x_{0}=1$, $x=3$, and $N=31$ sites. Exact numerical results are obtained for $v/w=1$ (blue circles), $v/w=0.8$ (red squares), $v/w=0.6$ (green diamonds) by performing the evolution up to time $5\times 10^6 w^{-1}$ in steps of $100 w^{-1}$. The solid lines stand for the corresponding noncentral chi-squared distributions [Eq. (\ref{NC})].
Note that at the metallic critical point $v/w=1$, the distribution approaches the standard chi-squared (exponential) in the absence (presence) of the disorder. In terms of the physical scheme in (a,b) it means that $C_{\mathrm{e}}$ becomes obscured by the noise.  
}
\end{figure}

In the absence of on-site disorder the SSH Hamiltonian preserves chiral symmetry, which implies $E_{k}=-E_{-k}$ and $C_{k}=\pm C_{-k}$. 
Then, considering odd values of $x_0$ and $x$ so that $C_{\mathrm{e}}$ does not trivially vanish,
Eq. (\ref{phasor}) assumes $\Psi_x=C_{\mathrm{e}}+\sum_{k'} 2 C_{k'} \cos (\phi_{k'})$, with the sum over $k'$ running one half of the $M$ modes (the time dependence of $\Psi_x$ and the phases is hereby omitted for brevity). 
By resorting to the central limit theorem we argue that many realizations of $\Psi_x$ (via the truncated time evolution) result in a constant phasor $C_{\mathrm{e}}$ perturbed by a Gaussian noise with zero mean and variance $\sigma^2=2\sum_{k'}C_{k'}^2$ on the real axis, as depicted in Fig. \ref{fig2}(a).
%
%
Hence, $\Psi_x$ is also normally distributed 
with a shifted mean $C_{\mathrm{e}}$. In turn, the amplitude $A=|\Psi_x|$ obeys a folded normal distribution. 

In that scenario the intensity $I= A^2/\sigma^{2}$ is distributed according to the noncentral chi-squared distribution
\begin{equation}\label{NC}
p_I(I;\mu,\lambda)=\frac{1}{2}e^{-(I+\lambda)/2}\left(\frac{I}{\lambda}\right)^{\mu/4-1/2}\mathcal{I}_{\mu/2-1}(\sqrt{\lambda I}),
\end{equation}
 with one degree of freedom ($\mu=1$), noncentrality parameter $\lambda=(C_{\mathrm{e}}/\sigma)^2$, 
where $\mathcal{I}_\alpha (z)$ is the modified Bessel function of the first kind.
The distribution above agrees well with the numerical data shown in Fig. \ref{fig2}(c). 


In the presence of 
on-site disorder, the chiral symmetry is broken. So let us now consider that 
$\epsilon_x/w$ 
is a random variable uniformly distributed in the interval 
$[-0.01,0.01]$.
Given such a weak disorder, no substantial changes to the spectrum take place. The energy associated to the edge state $E_{\mathrm{e}}$ will slightly deviate from the center of the band, with $\psi_{\mathrm{e},x}$ (and $C_{\mathrm{e}}$)
maintaining its form as in
Eq. (\ref{bound}) [and Fig. \ref{fig1}(d)] for all practical purposes.  

The time evolution in the disordered case reads 
$\Psi_x=C_{\mathrm{e}}e^{-i\phi_{\mathrm{e}}}+\sum_kC_ke^{-i\phi_{k}}$. A convenient global phase factor $e^{i\phi_{\mathrm{e}}}$ can be set up in order to leave $C_{\mathrm{e}}$ as a constant phasor lying at the real axis for convenience.   
We readily see that  
the second term now describes
 a \textit{circular} Gaussian noise [see Fig. \ref{fig2}(b)] with the variance of both real and imaginary parts $\sigma'^2=(1/2)\sum_k C_{k}^2$ as a result of the broken chiral symmetry. 
The intensity $I= A^2/\sigma'^{2}$ is, again, described by the noncentral chi-squared distribution [Eq. (\ref{NC})], with noncentrality parameter $\lambda \rightarrow \lambda'=(C_{\mathrm{e}}/\sigma')^2$, now with two degrees of freedom $\mu=2$.
Note that when $\lambda' = 0$ the distribution becomes exponential, which is the standard intensity speckle regime. 
%
Figure \ref{fig2}(d)
compares numerical data with 
the expected probability density function. As in Fig. \ref{fig2}(c) we note that the dimerization of the chain contributes to the tail retraction, rendering a sub-Rayleigh speckle regime. We will see shortly that such a response 
has a nontrivial dependence 
on $x_{0}$ and $x$. 

The noncentrality parameters $\lambda$ and $\lambda'$ ultimately define the shape of the speckle distribution in the clean and disordered cases, respectively.
An analytical expression for them is in order.
First, we assume that the edge-state phasor amplitude $C_{\mathrm{e}}$ is the same in both situations. 
%
The variance $\sigma^2$ for the ordered case
can be approximated by making \footnote{It is a reasonable approximation given we are only interested in the statistics of the local wavefunction amplitude, rather than its exact time evolution.} $\psi_{k,x}\approx [(1-|\psi_{\mathrm{e},x}|^2)/M]^{1/2}$ for all $k$ and thus $\sigma^2=MC_{k}^{2}$. We therefore obtain
\begin{equation}\label{lambda}
\lambda=\frac{C_{\mathrm{e}}^2}{MC_k^2}=\frac{v^{2(m_0+m)}(v^2-1)^2 M}{\tilde{v}_{m_0}\tilde{v}_{m}},
\end{equation}
where
$\tilde{v}_{m}=v^2-v^{M+4}-v^{2m}+v^{2+2m}$ and $m=(x+1)/2$.
A similar reasoning can be made for the disordered case, despite the fact that the chiral symmetry is broken, to obtain $\sigma'^2=\sigma^2 / 2$ and
$\lambda' = 2\lambda$.

\begin{figure}[t!] 
\includegraphics[width=0.45\textwidth]{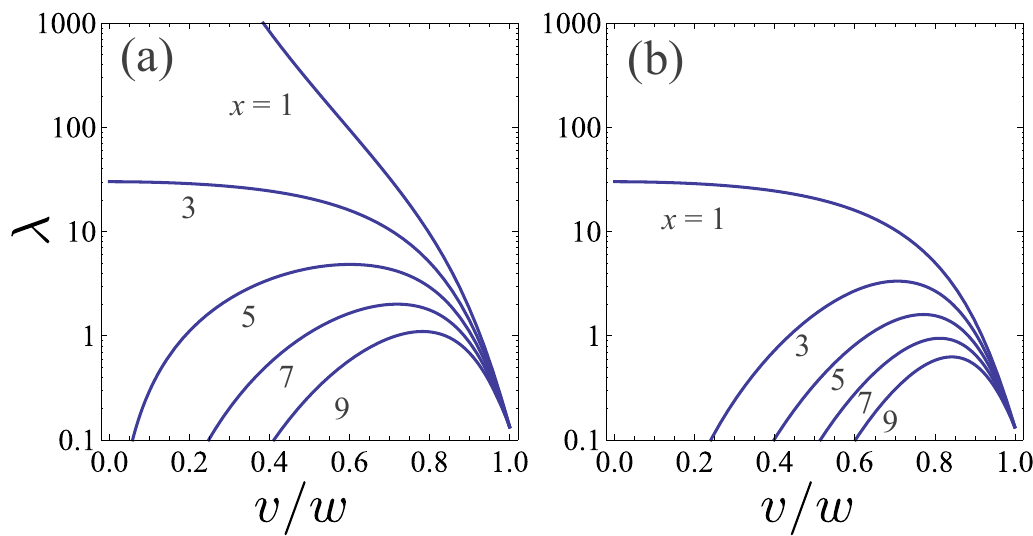}
\caption{\label{fig3} 
Noncentrality parameter $\lambda$ 
evaluated from Eq. (\ref{lambda}) against the distortion of the chain given by $v/w$ ($w\equiv 1$) for $N=31$, and input locations (a) $x_{0}=1$ and (b) $x_{0}=3$. Output 
locations are shown on each panel. 
}
\end{figure}

The behavior of $\lambda$ against the dimerization parameter $v/w$ is shown in Fig. \ref{fig3}. As expected, $\lambda$ always
approaches zero as the chain becomes homogeneous ($v/w\rightarrow 1$) meaning that $p_I(I)$ takes the form of the standard chi-squared distribution (exponential distribution) in the absence (presence) of disorder.
However,  
apart from a few cases in Fig. \ref{fig3}, $\lambda$ rebounds to zero as the chain undergoes dimerization. This can be understood by realizing that $\lambda$ measures how much the constant phasor $C_{\mathrm{e}}(x_0,x)$ is blurred 
into the Gaussian noise [second term of Eq. (\ref{phasor})]. 
The farther from the main edge one chooses $x_0$ and $x$ the Gaussian noise rapidly dominates as $v/w \rightarrow 0$ .
%

Note that in a homogeneous chain with an odd number of sites the zero-mode wavefunction is 
$\propto [(N+1)/2]^{-1/2}$ for all odd $x$.
As soon as $v/w < 1$ the edge state
emerges with its exponential fall-off across the bulk. 
Given $x\neq 1$ and not too far from the main edge, the wavefunction amplitude will peak at some threshold value of $v/w$ before vanishing so as to conform with 
$|\psi_{\mathrm{e},1}|\rightarrow 1$ as $v/w\rightarrow 0$.
That is, in the dimerized limit the edge state becomes fully localized at the first site, what explains
the divergence of $\lambda$ when $(x_{0},x)=(1,1)$.
Another particular behavior is observed for 
$(x_{0},x)=(1,3)$ in Fig. \ref{fig3}.
The saturation of $\lambda$ as $v/w\rightarrow 0$ in this case points out a similar functional dependence of the decay 
rates of $C_{\mathrm{e}}^2$ and $C_{\mathrm{k}}^2$. Indeed, they both decay $\sim (v/w)^2$ at that limit.  
%

As one may have realized at this point, the speckle 
contrast -- standard deviation of the intensity over its mean -- $K=\sigma_{I}/\langle I \rangle$ is a function of $\lambda$.
With respect to the noncentral chi-squared distribution in Eq. (\ref{NC}) it assumes $K=\sqrt{2(\mu+2\lambda)}/(\mu+\lambda)$. Thus, notice that 
the functions corresponding 
to clean ($\mu=1$) and disordered ($\mu=2$) cases differ by a factor of $\sqrt{2}$ considering the expression for $\lambda$ in Eq. (\ref{lambda}), with $\lambda \rightarrow \lambda'=2\lambda$ for the latter case. 

\begin{figure*}[t!] 
\includegraphics[width=0.9\textwidth]{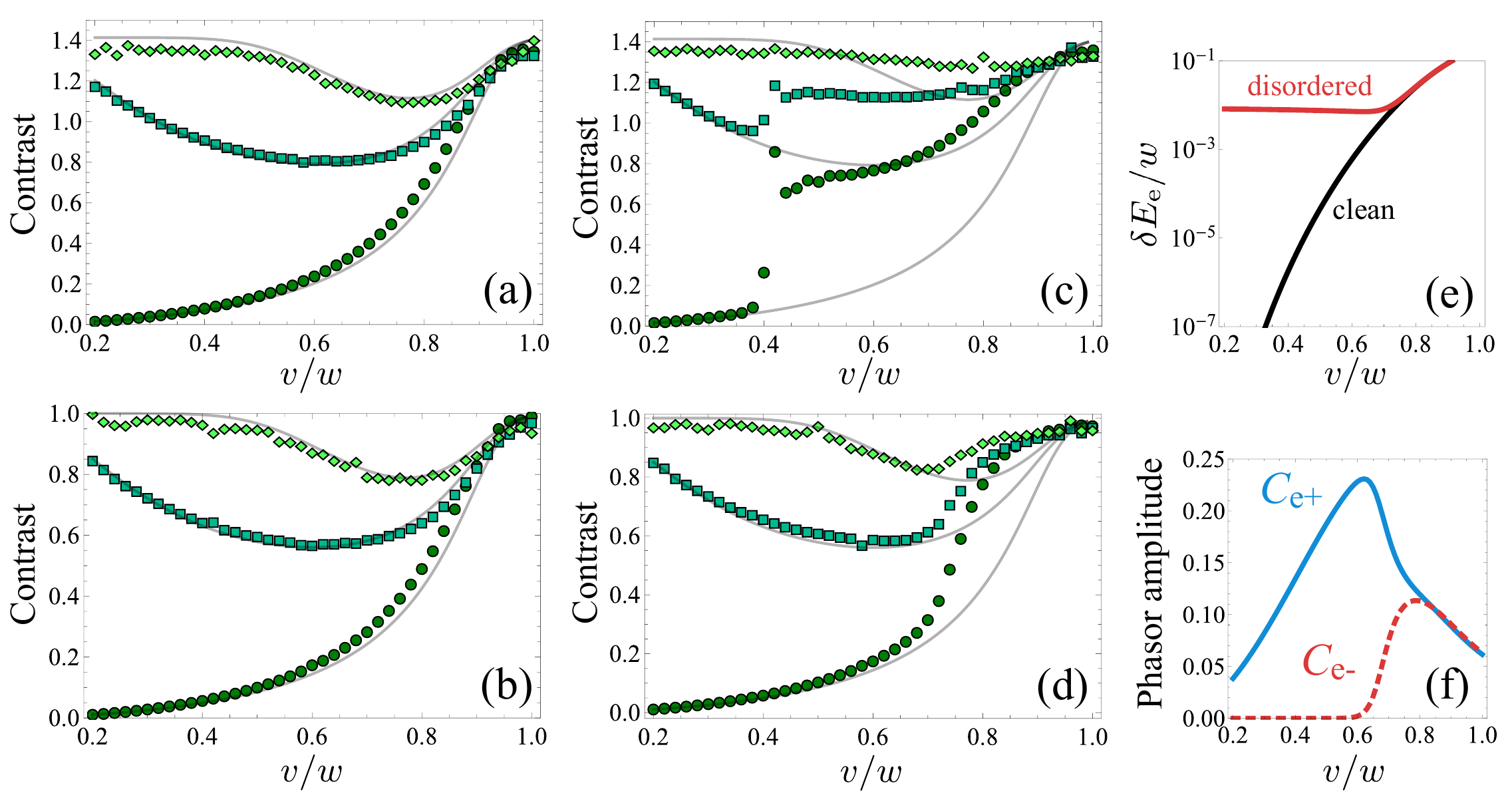}
\caption{\label{fig4}
(a,c) Speckle contrasts versus  $v/w$ for the clean and (b,d) disordered SSH chains, with $N=31$ (first column) and $N=30$ (second column).
Input and output sites $(x_0,x)$ are: (1,1) [circles]; (1,5) [squares]; (3,5) [diamonds]. 
In all cases the unitary
evolution is carried out up to time $10^6 w^{-1}$ in steps of $100 w^{-1}$. Only one disordered sample is considered, where the on-site energies are uniformly distributed within $[-0.01w,0.01w]$.  
Gray solid lines represent
the analytical contrasts
$K=\sqrt{2+4\lambda}/(1+\lambda)$ and $K=\sqrt{1+2\lambda}/(1+\lambda)$ for the clean and disordered cases, respectively, with
the parameter $\lambda$ being calculated from Eq. (\ref{lambda}), valid for odd $N$. Eventually, for low enough $v/w$, the speckle behavior corresponding to even $N$ approaches that of odd $N$, which features a single edge state. In the clean case, such an information loss regarding the existence of a pair of edge states when $N$ is even occurs because the closing energy gap $\delta E_{\mathrm{e}}$ in the middle of the band as $v/w \rightarrow 0$, as shown in panel (e). 
A dynamical transition from even- to odd-$N$ behavior also occurs
in the disordered case, but it is  not driven by the closing gap. Rather, the pair of edges states break apart due to the disorder, forcing one of the dominant phasors to vanish, as displayed in panel (f), with $C_{\mathrm{e}\pm}$ evaluated for $(x_0=1,x=5)$. 
}
\end{figure*}

Figures \ref{fig4}(a) and \ref{fig4}(b) show $K$ versus $v/w$ for different locations in the chain.
We confirm that the analytical formula fits well to the exact numerical results. Note that as the chain becomes homogeneous the contrast converges to unit ($\sqrt{2}$) in the presence (absence) of the disorder, 
which relates to the standard exponential (chi-squared) speckle statistics. 
The same regime can be achieved 
at a finite dimerization ratio as soon as $x_0$ and $x$ are picked a few sites away from the edge 
(follow the diamond symbols in the figures and cf. Fig. \ref{fig3}).



\textit{Bilocalized edge states.}
Let us now move on to the speckle properties when the SSH chain hosts two edge states. 
Setting an even $N$, with $v/w<1$ (nontrivial topological phase), we induce the formation of a pair of edge states $\psi_{\mathrm{e\pm}}$,
with energies $E_{\mathrm{e\pm}}$,
exponentially decaying from both edges [as previously showed in Fig. \ref{fig1}(b)]. 
We will not worry here about their analytical form but it suffices to mention that their wavefunction is almost a mirrored version of Eq. (\ref{bound}) with proper symmetrization. A solution based on perturbation theory can be found in Ref. \cite{almeida16}. 

For a symmetric spectrum, the time-evolved phasor reads $\Psi_x=2C_{\mathrm{e\pm}}\cos(\phi_{\delta})+\sum_{k'}2C_{k'}\cos(\phi_{k'})$, 
where $k'$ covers $(N-2)/2$ delocalized modes, $C_{\mathrm{e\pm}} = C_{\mathrm{e+}}=C_{\mathrm{e-}}$, and $\phi_{\delta}=\delta E_{\mathrm{e}}t/2$.
The above encompasses a situation similar to the one depicted in Fig. \ref{fig2}(a) except that the dominant phasor is no longer constant. It evolves periodically
at a rate that depends 
on the gap $\delta E_{\mathrm{e}}$. 

In principle, 
the distribution for $I$ can be obtained by compounding the noncentral chi-squared distribution (with degree $\mu=1$), Eq. (\ref{NC}), over various $\lambda$. 
Note, however, 
that the gap at the middle of the band obeys $\delta E_{\mathrm{e}}\sim (v/w)^{N/2+1}$ \cite{almeida18}, as stated earlier. 
So depending on the considered time window the cosine function may stall, rendering a constant edge-state phasor.  
In Fig. \ref{fig4}(c) we plot the speckle contrast against $v/w$ and confirm such a behavior by the moment $\delta E_{\mathrm{e}}$ reaches about $10^{-6}w$, that is of the order of the inverse of the maximum evolution time [see Fig. \ref{fig4}(e)].  
Note that contrast maintains an upper bound of $\sqrt{2}$ despite the value of $v/w$.

In the disordered SSH chain
with an even $N$
the evolved wavefunction can be cast in the form
$\Psi_x = C_{\mathrm{e+}}+C_{\mathrm{e-}}e^{i\phi_\delta}+\sum_k C_ke^{i(\phi_{\mathrm{e}+}-\phi_{k})}$, with the last term comprising the $N-2$ delocalized modes representing the circular Gaussian noise. 
This time there is a couple of dominant phasors, one of which rotates at a rate defined by the gap $\delta E_{\mathrm{e}}$.
Differently from the previous regime, though, the gap will not vanish due to the disorder [see Fig. \ref{fig4}(e)]. On the other hand, one of the edge-state phasors will die off [an example is shown in \ref{fig4}(f)]
as the on-site disorder destroys their mirror symmetry. Indeed, each state acquire a form similar to those depicted in Fig. \ref{fig1}(d).

Now the contrast responds to the dimerization
as shown in Fig. \ref{fig4}(d).
The intensity fluctuations eventually develop
as in the odd-$N$ case, where 
 a constant, single edge-state phasor is surrounded by a circular Gaussian cloud [see Fig. \ref{fig2}(b)].  
The transition between dynamical regimes is smoother compared to the one displayed in Fig. \ref{fig4}(c) for the ordered case, as they are driven by distinct sources. 


\textit{Conclusion and outlook.}
We thereby realize that local measurements of the wavefunction
at distinct times
reveal intrinsic topological properties of the lattice. For a range of  $v/w$ values 
the intensity speckles can be used to detect the presence of symmetry-breaking disorder, the gap at the middle of the band, and the degree of edge-state localization. Thus,
the mapping of the natural Hamiltonian dynamics of the system onto random phasor models offers a route for characterization of involved dynamical regimes. 
It can even be used to measure
the degree of quantum correlations \cite{beenakker09, oliveira23}. 
For instance, if the SSH dimerized pattern of couplings were applied to a isotropic $XY$ spin$-1/2$ chain \cite{venuti07}, the quantity $2|C_{k}(x_0,x)|$ would be the concurrence between the spins in those positions with respect to the mode $k$ \cite{amico04}. In that scenario, 
lower contrasts would indicate a higher degree of bipartite entanglement present in the edge-state.

%

The study of speckle statistics in topological lattices may also shed new light on the nature of unconventional quantum states of matter.
Prototype physical systems to explore this venue would consist of staggered antiferromagnetic and ferrimagnetic quantum spin chains with $S>1/2$ spin units. These can depict topological ground states with unconventional entanglement features and even a sequence of gap-closing phase transitions involving edge states with distinct spin configurations \cite{haldane83, oshikawa97, verissimo21, xu03}. 

The SSH model is feasible for an experimental implementation based on the propagation of classical light through linear coupled waveguide lattices \cite{kang23}.
The staggered hopping constant is realized by the spacing between waveguides and the Hamiltonian time evolution is analog to the beam propagation along the longitudinal axis. Other potential physical platforms for realizing topological lattices 
are in order such as arrays of trapped ions \cite{nevado17} and superconducting circuits \cite{youssefi22}.

In terms of speckle customization, a natural extension of our work should include the role of 
coupling disorder, which preserves the chiral symmetry. An interesting variation of the model is based on a set of interacting SSH moduli \cite{almeida16,zurita23}. This kind of arrangement can be used to control the size of the gap at the middle of the band \cite{almeida16}.  

The quest for achieving high-precision tailored intensity statistics goes far beyond meeting
practical applications. The ability to extract information from noise
is paramount to achieve a deeper understanding of nature.
We hope that our results fuel further research on both ends. 

%
%




This work was supported by CNPq, CAPES (Brazilian agencies),
and FAPEAL (Alagoas state agency). 


\begin{thebibliography}{34}%
\makeatletter
\providecommand \@ifxundefined [1]{%
 \@ifx{#1\undefined}
}%
\providecommand \@ifnum [1]{%
 \ifnum #1\expandafter \@firstoftwo
 \else \expandafter \@secondoftwo
 \fi
}%
\providecommand \@ifx [1]{%
 \ifx #1\expandafter \@firstoftwo
 \else \expandafter \@secondoftwo
 \fi
}%
\providecommand \natexlab [1]{#1}%
\providecommand \enquote  [1]{``#1''}%
\providecommand \bibnamefont  [1]{#1}%
\providecommand \bibfnamefont [1]{#1}%
\providecommand \citenamefont [1]{#1}%
\providecommand \href@noop [0]{\@secondoftwo}%
\providecommand \href [0]{\begingroup \@sanitize@url \@href}%
\providecommand \@href[1]{\@@startlink{#1}\@@href}%
\providecommand \@@href[1]{\endgroup#1\@@endlink}%
\providecommand \@sanitize@url [0]{\catcode `\\12\catcode `\$12\catcode `\&12\catcode `\#12\catcode `\^12\catcode `\_12\catcode `\%12\relax}%
\providecommand \@@startlink[1]{}%
\providecommand \@@endlink[0]{}%
\providecommand \url  [0]{\begingroup\@sanitize@url \@url }%
\providecommand \@url [1]{\endgroup\@href {#1}{\urlprefix }}%
\providecommand \urlprefix  [0]{URL }%
\providecommand \Eprint [0]{\href }%
\providecommand \doibase [0]{https://doi.org/}%
\providecommand \selectlanguage [0]{\@gobble}%
\providecommand \bibinfo  [0]{\@secondoftwo}%
\providecommand \bibfield  [0]{\@secondoftwo}%
\providecommand \translation [1]{[#1]}%
\providecommand \BibitemOpen [0]{}%
\providecommand \bibitemStop [0]{}%
\providecommand \bibitemNoStop [0]{.\EOS\space}%
\providecommand \EOS [0]{\spacefactor3000\relax}%
\providecommand \BibitemShut  [1]{\csname bibitem#1\endcsname}%
\let\auto@bib@innerbib\@empty
\bibitem [{\citenamefont {Goodman}(2020)}]{goodmanbook}%
  \BibitemOpen
  \bibfield  {author} {\bibinfo {author} {\bibfnamefont {J.}~\bibnamefont {Goodman}},\ }\href {https://books.google.com.br/books?id=3FvmyAEACAAJ} {\emph {\bibinfo {title} {Speckle Phenomena in Optics: Theory and Applications}}},\ Press Monographs\ (\bibinfo  {publisher} {SPIE Press},\ \bibinfo {year} {2020})\BibitemShut {NoStop}%
\bibitem [{\citenamefont {Heeman}\ \emph {et~al.}(2019)\citenamefont {Heeman}, \citenamefont {Steenbergen}, \citenamefont {van Dam},\ and\ \citenamefont {Boerma}}]{heeman19}%
  \BibitemOpen
  \bibfield  {author} {\bibinfo {author} {\bibfnamefont {W.}~\bibnamefont {Heeman}}, \bibinfo {author} {\bibfnamefont {W.}~\bibnamefont {Steenbergen}}, \bibinfo {author} {\bibfnamefont {G.~M.}\ \bibnamefont {van Dam}},\ and\ \bibinfo {author} {\bibfnamefont {E.~C.}\ \bibnamefont {Boerma}},\ }\bibfield  {title} {\bibinfo {title} {Clinical applications of laser speckle contrast imaging: a review},\ }\href {https://doi.org/10.1117/1.JBO.24.8.080901} {\bibfield  {journal} {\bibinfo  {journal} {Journal of Biomedical Optics}\ }\textbf {\bibinfo {volume} {24}},\ \bibinfo {pages} {080901} (\bibinfo {year} {2019})}\BibitemShut {NoStop}%
\bibitem [{\citenamefont {Verwohlt}\ \emph {et~al.}(2018)\citenamefont {Verwohlt}, \citenamefont {Reiser}, \citenamefont {Randolph}, \citenamefont {Matic}, \citenamefont {Medina}, \citenamefont {Madsen}, \citenamefont {Sprung}, \citenamefont {Zozulya},\ and\ \citenamefont {Gutt}}]{verwohlt18}%
  \BibitemOpen
  \bibfield  {author} {\bibinfo {author} {\bibfnamefont {J.}~\bibnamefont {Verwohlt}}, \bibinfo {author} {\bibfnamefont {M.}~\bibnamefont {Reiser}}, \bibinfo {author} {\bibfnamefont {L.}~\bibnamefont {Randolph}}, \bibinfo {author} {\bibfnamefont {A.}~\bibnamefont {Matic}}, \bibinfo {author} {\bibfnamefont {L.~A.}\ \bibnamefont {Medina}}, \bibinfo {author} {\bibfnamefont {A.}~\bibnamefont {Madsen}}, \bibinfo {author} {\bibfnamefont {M.}~\bibnamefont {Sprung}}, \bibinfo {author} {\bibfnamefont {A.}~\bibnamefont {Zozulya}},\ and\ \bibinfo {author} {\bibfnamefont {C.}~\bibnamefont {Gutt}},\ }\bibfield  {title} {\bibinfo {title} {Low dose x-ray speckle visibility spectroscopy reveals nanoscale dynamics in radiation sensitive ionic liquids},\ }\href {https://doi.org/10.1103/PhysRevLett.120.168001} {\bibfield  {journal} {\bibinfo  {journal} {Phys. Rev. Lett.}\ }\textbf {\bibinfo {volume} {120}},\ \bibinfo {pages} {168001} (\bibinfo {year} {2018})}\BibitemShut {NoStop}%
\bibitem [{\citenamefont {Zhang}\ \emph {et~al.}(2023)\citenamefont {Zhang}, \citenamefont {Gamekkanda}, \citenamefont {Pandit}, \citenamefont {Tang}, \citenamefont {Papageorgiou}, \citenamefont {Mitchell}, \citenamefont {Yang}, \citenamefont {Schwaerzler}, \citenamefont {Oyetunde}, \citenamefont {Braatz}, \citenamefont {Myerson},\ and\ \citenamefont {Barbastathis}}]{zhang23}%
  \BibitemOpen
  \bibfield  {author} {\bibinfo {author} {\bibfnamefont {Q.}~\bibnamefont {Zhang}}, \bibinfo {author} {\bibfnamefont {J.~C.}\ \bibnamefont {Gamekkanda}}, \bibinfo {author} {\bibfnamefont {A.}~\bibnamefont {Pandit}}, \bibinfo {author} {\bibfnamefont {W.}~\bibnamefont {Tang}}, \bibinfo {author} {\bibfnamefont {C.}~\bibnamefont {Papageorgiou}}, \bibinfo {author} {\bibfnamefont {C.}~\bibnamefont {Mitchell}}, \bibinfo {author} {\bibfnamefont {Y.}~\bibnamefont {Yang}}, \bibinfo {author} {\bibfnamefont {M.}~\bibnamefont {Schwaerzler}}, \bibinfo {author} {\bibfnamefont {T.}~\bibnamefont {Oyetunde}}, \bibinfo {author} {\bibfnamefont {R.~D.}\ \bibnamefont {Braatz}}, \bibinfo {author} {\bibfnamefont {A.~S.}\ \bibnamefont {Myerson}},\ and\ \bibinfo {author} {\bibfnamefont {G.}~\bibnamefont {Barbastathis}},\ }\bibfield  {title} {\bibinfo {title} {Extracting particle size distribution from laser speckle with a physics-enhanced autocorrelation-based estimator {(PEACE)}},\ }\href {https://doi.org/10.1038/s41467-023-36816-2}
  {\bibfield  {journal} {\bibinfo  {journal} {Nature Communications}\ }\textbf {\bibinfo {volume} {14}},\ \bibinfo {pages} {1159} (\bibinfo {year} {2023})}\BibitemShut {NoStop}%
\bibitem [{\citenamefont {Luo}\ \emph {et~al.}(2021)\citenamefont {Luo}, \citenamefont {Patel},\ and\ \citenamefont {Webb}}]{luo21}%
  \BibitemOpen
  \bibfield  {author} {\bibinfo {author} {\bibfnamefont {Q.}~\bibnamefont {Luo}}, \bibinfo {author} {\bibfnamefont {J.~A.}\ \bibnamefont {Patel}},\ and\ \bibinfo {author} {\bibfnamefont {K.~J.}\ \bibnamefont {Webb}},\ }\bibfield  {title} {\bibinfo {title} {Super-resolution sensing with a randomly scattering analyzer},\ }\href {https://doi.org/10.1103/PhysRevResearch.3.L042045} {\bibfield  {journal} {\bibinfo  {journal} {Phys. Rev. Res.}\ }\textbf {\bibinfo {volume} {3}},\ \bibinfo {pages} {L042045} (\bibinfo {year} {2021})}\BibitemShut {NoStop}%
\bibitem [{\citenamefont {Jendrzejewski}\ \emph {et~al.}(2012)\citenamefont {Jendrzejewski}, \citenamefont {Bernard}, \citenamefont {M{\"u}ller}, \citenamefont {Cheinet}, \citenamefont {Josse}, \citenamefont {Piraud}, \citenamefont {Pezz{\'e}}, \citenamefont {Sanchez-Palencia}, \citenamefont {Aspect},\ and\ \citenamefont {Bouyer}}]{jendrzejewski12}%
  \BibitemOpen
  \bibfield  {author} {\bibinfo {author} {\bibfnamefont {F.}~\bibnamefont {Jendrzejewski}}, \bibinfo {author} {\bibfnamefont {A.}~\bibnamefont {Bernard}}, \bibinfo {author} {\bibfnamefont {K.}~\bibnamefont {M{\"u}ller}}, \bibinfo {author} {\bibfnamefont {P.}~\bibnamefont {Cheinet}}, \bibinfo {author} {\bibfnamefont {V.}~\bibnamefont {Josse}}, \bibinfo {author} {\bibfnamefont {M.}~\bibnamefont {Piraud}}, \bibinfo {author} {\bibfnamefont {L.}~\bibnamefont {Pezz{\'e}}}, \bibinfo {author} {\bibfnamefont {L.}~\bibnamefont {Sanchez-Palencia}}, \bibinfo {author} {\bibfnamefont {A.}~\bibnamefont {Aspect}},\ and\ \bibinfo {author} {\bibfnamefont {P.}~\bibnamefont {Bouyer}},\ }\bibfield  {title} {\bibinfo {title} {Three-dimensional localization of ultracold atoms in an optical disordered potential},\ }\href {https://doi.org/10.1038/nphys2256} {\bibfield  {journal} {\bibinfo  {journal} {Nature Physics}\ }\textbf {\bibinfo {volume} {8}},\ \bibinfo {pages} {398} (\bibinfo {year} {2012})}\BibitemShut {NoStop}%
\bibitem [{\citenamefont {Delande}\ and\ \citenamefont {Orso}(2014)}]{delande14}%
  \BibitemOpen
  \bibfield  {author} {\bibinfo {author} {\bibfnamefont {D.}~\bibnamefont {Delande}}\ and\ \bibinfo {author} {\bibfnamefont {G.}~\bibnamefont {Orso}},\ }\bibfield  {title} {\bibinfo {title} {Mobility edge for cold atoms in laser speckle potentials},\ }\href {https://doi.org/10.1103/PhysRevLett.113.060601} {\bibfield  {journal} {\bibinfo  {journal} {Phys. Rev. Lett.}\ }\textbf {\bibinfo {volume} {113}},\ \bibinfo {pages} {060601} (\bibinfo {year} {2014})}\BibitemShut {NoStop}%
\bibitem [{\citenamefont {Bromberg}\ and\ \citenamefont {Cao}(2014)}]{bromberg14}%
  \BibitemOpen
  \bibfield  {author} {\bibinfo {author} {\bibfnamefont {Y.}~\bibnamefont {Bromberg}}\ and\ \bibinfo {author} {\bibfnamefont {H.}~\bibnamefont {Cao}},\ }\bibfield  {title} {\bibinfo {title} {Generating non-{R}ayleigh speckles with tailored intensity statistics},\ }\href {https://doi.org/10.1103/PhysRevLett.112.213904} {\bibfield  {journal} {\bibinfo  {journal} {Phys. Rev. Lett.}\ }\textbf {\bibinfo {volume} {112}},\ \bibinfo {pages} {213904} (\bibinfo {year} {2014})}\BibitemShut {NoStop}%
\bibitem [{\citenamefont {Han}\ \emph {et~al.}(2023)\citenamefont {Han}, \citenamefont {Bender},\ and\ \citenamefont {Cao}}]{han23}%
  \BibitemOpen
  \bibfield  {author} {\bibinfo {author} {\bibfnamefont {S.}~\bibnamefont {Han}}, \bibinfo {author} {\bibfnamefont {N.}~\bibnamefont {Bender}},\ and\ \bibinfo {author} {\bibfnamefont {H.}~\bibnamefont {Cao}},\ }\bibfield  {title} {\bibinfo {title} {Tailoring {3D} speckle statistics},\ }\href {https://doi.org/10.1103/PhysRevLett.130.093802} {\bibfield  {journal} {\bibinfo  {journal} {Phys. Rev. Lett.}\ }\textbf {\bibinfo {volume} {130}},\ \bibinfo {pages} {093802} (\bibinfo {year} {2023})}\BibitemShut {NoStop}%
\bibitem [{\citenamefont {Bender}\ \emph {et~al.}(2023)\citenamefont {Bender}, \citenamefont {Haig}, \citenamefont {Christodoulides},\ and\ \citenamefont {Wise}}]{bender23}%
  \BibitemOpen
  \bibfield  {author} {\bibinfo {author} {\bibfnamefont {N.}~\bibnamefont {Bender}}, \bibinfo {author} {\bibfnamefont {H.}~\bibnamefont {Haig}}, \bibinfo {author} {\bibfnamefont {D.~N.}\ \bibnamefont {Christodoulides}},\ and\ \bibinfo {author} {\bibfnamefont {F.~W.}\ \bibnamefont {Wise}},\ }\bibfield  {title} {\bibinfo {title} {Spectral speckle customization},\ }\href {https://doi.org/10.1364/OPTICA.499461} {\bibfield  {journal} {\bibinfo  {journal} {Optica}\ }\textbf {\bibinfo {volume} {10}},\ \bibinfo {pages} {1260} (\bibinfo {year} {2023})}\BibitemShut {NoStop}%
\bibitem [{\citenamefont {Oliveira}\ \emph {et~al.}(2023)\citenamefont {Oliveira}, \citenamefont {de~Moura}, \citenamefont {Souza}, \citenamefont {Lyra},\ and\ \citenamefont {Almeida}}]{oliveira23}%
  \BibitemOpen
  \bibfield  {author} {\bibinfo {author} {\bibfnamefont {M.~F.~V.}\ \bibnamefont {Oliveira}}, \bibinfo {author} {\bibfnamefont {F.~A. B.~F.}\ \bibnamefont {de~Moura}}, \bibinfo {author} {\bibfnamefont {A.~M.~C.}\ \bibnamefont {Souza}}, \bibinfo {author} {\bibfnamefont {M.~L.}\ \bibnamefont {Lyra}},\ and\ \bibinfo {author} {\bibfnamefont {G.~M.~A.}\ \bibnamefont {Almeida}},\ }\bibfield  {title} {\bibinfo {title} {Non-{R}ayleigh signal of interacting quantum particles},\ }\href {https://doi.org/10.1103/PhysRevA.108.023520} {\bibfield  {journal} {\bibinfo  {journal} {Phys. Rev. A}\ }\textbf {\bibinfo {volume} {108}},\ \bibinfo {pages} {023520} (\bibinfo {year} {2023})}\BibitemShut {NoStop}%
\bibitem [{\citenamefont {Beenakker}\ \emph {et~al.}(2009)\citenamefont {Beenakker}, \citenamefont {Venderbos},\ and\ \citenamefont {van Exter}}]{beenakker09}%
  \BibitemOpen
  \bibfield  {author} {\bibinfo {author} {\bibfnamefont {C.~W.~J.}\ \bibnamefont {Beenakker}}, \bibinfo {author} {\bibfnamefont {J.~W.~F.}\ \bibnamefont {Venderbos}},\ and\ \bibinfo {author} {\bibfnamefont {M.~P.}\ \bibnamefont {van Exter}},\ }\bibfield  {title} {\bibinfo {title} {Two-photon speckle as a probe of multi-dimensional entanglement},\ }\href {https://doi.org/10.1103/PhysRevLett.102.193601} {\bibfield  {journal} {\bibinfo  {journal} {Phys. Rev. Lett.}\ }\textbf {\bibinfo {volume} {102}},\ \bibinfo {pages} {193601} (\bibinfo {year} {2009})}\BibitemShut {NoStop}%
\bibitem [{\citenamefont {Peeters}\ \emph {et~al.}(2010)\citenamefont {Peeters}, \citenamefont {Moerman},\ and\ \citenamefont {van Exter}}]{peeters10}%
  \BibitemOpen
  \bibfield  {author} {\bibinfo {author} {\bibfnamefont {W.~H.}\ \bibnamefont {Peeters}}, \bibinfo {author} {\bibfnamefont {J.~J.~D.}\ \bibnamefont {Moerman}},\ and\ \bibinfo {author} {\bibfnamefont {M.~P.}\ \bibnamefont {van Exter}},\ }\bibfield  {title} {\bibinfo {title} {Observation of two-photon speckle patterns},\ }\href {https://doi.org/10.1103/PhysRevLett.104.173601} {\bibfield  {journal} {\bibinfo  {journal} {Phys. Rev. Lett.}\ }\textbf {\bibinfo {volume} {104}},\ \bibinfo {pages} {173601} (\bibinfo {year} {2010})}\BibitemShut {NoStop}%
\bibitem [{\citenamefont {Di~Lorenzo~Pires}\ \emph {et~al.}(2012)\citenamefont {Di~Lorenzo~Pires}, \citenamefont {Woudenberg},\ and\ \citenamefont {van Exter}}]{pires12}%
  \BibitemOpen
  \bibfield  {author} {\bibinfo {author} {\bibfnamefont {H.}~\bibnamefont {Di~Lorenzo~Pires}}, \bibinfo {author} {\bibfnamefont {J.}~\bibnamefont {Woudenberg}},\ and\ \bibinfo {author} {\bibfnamefont {M.~P.}\ \bibnamefont {van Exter}},\ }\bibfield  {title} {\bibinfo {title} {Statistical properties of two-photon speckles},\ }\href {https://doi.org/10.1103/PhysRevA.85.033807} {\bibfield  {journal} {\bibinfo  {journal} {Phys. Rev. A}\ }\textbf {\bibinfo {volume} {85}},\ \bibinfo {pages} {033807} (\bibinfo {year} {2012})}\BibitemShut {NoStop}%
\bibitem [{\citenamefont {Kirkby}\ \emph {et~al.}(2022)\citenamefont {Kirkby}, \citenamefont {Yee}, \citenamefont {Shi},\ and\ \citenamefont {O'Dell}}]{kirkby22}%
  \BibitemOpen
  \bibfield  {author} {\bibinfo {author} {\bibfnamefont {W.}~\bibnamefont {Kirkby}}, \bibinfo {author} {\bibfnamefont {Y.}~\bibnamefont {Yee}}, \bibinfo {author} {\bibfnamefont {K.}~\bibnamefont {Shi}},\ and\ \bibinfo {author} {\bibfnamefont {D.~H.~J.}\ \bibnamefont {O'Dell}},\ }\bibfield  {title} {\bibinfo {title} {Caustics in quantum many-body dynamics},\ }\href {https://doi.org/10.1103/PhysRevResearch.4.013105} {\bibfield  {journal} {\bibinfo  {journal} {Phys. Rev. Research}\ }\textbf {\bibinfo {volume} {4}},\ \bibinfo {pages} {013105} (\bibinfo {year} {2022})}\BibitemShut {NoStop}%
\bibitem [{\citenamefont {Buarque}\ \emph {et~al.}(2022)\citenamefont {Buarque}, \citenamefont {Dias}, \citenamefont {de~Moura}, \citenamefont {Lyra},\ and\ \citenamefont {Almeida}}]{buarque22}%
  \BibitemOpen
  \bibfield  {author} {\bibinfo {author} {\bibfnamefont {A.~R.~C.}\ \bibnamefont {Buarque}}, \bibinfo {author} {\bibfnamefont {W.~S.}\ \bibnamefont {Dias}}, \bibinfo {author} {\bibfnamefont {F.~A. B.~F.}\ \bibnamefont {de~Moura}}, \bibinfo {author} {\bibfnamefont {M.~L.}\ \bibnamefont {Lyra}},\ and\ \bibinfo {author} {\bibfnamefont {G.~M.~A.}\ \bibnamefont {Almeida}},\ }\bibfield  {title} {\bibinfo {title} {Rogue waves in discrete-time quantum walks},\ }\href {https://doi.org/10.1103/PhysRevA.106.012414} {\bibfield  {journal} {\bibinfo  {journal} {Phys. Rev. A}\ }\textbf {\bibinfo {volume} {106}},\ \bibinfo {pages} {012414} (\bibinfo {year} {2022})}\BibitemShut {NoStop}%
\bibitem [{\citenamefont {Buarque}\ \emph {et~al.}(2023)\citenamefont {Buarque}, \citenamefont {Dias}, \citenamefont {Almeida}, \citenamefont {Lyra},\ and\ \citenamefont {de~Moura}}]{buarque22-2}%
  \BibitemOpen
  \bibfield  {author} {\bibinfo {author} {\bibfnamefont {A.~R.~C.}\ \bibnamefont {Buarque}}, \bibinfo {author} {\bibfnamefont {W.~S.}\ \bibnamefont {Dias}}, \bibinfo {author} {\bibfnamefont {G.~M.~A.}\ \bibnamefont {Almeida}}, \bibinfo {author} {\bibfnamefont {M.~L.}\ \bibnamefont {Lyra}},\ and\ \bibinfo {author} {\bibfnamefont {F.~A. B.~F.}\ \bibnamefont {de~Moura}},\ }\bibfield  {title} {\bibinfo {title} {Rogue waves in quantum lattices with correlated disorder},\ }\href {https://doi.org/10.1103/PhysRevA.107.012425} {\bibfield  {journal} {\bibinfo  {journal} {Phys. Rev. A}\ }\textbf {\bibinfo {volume} {107}},\ \bibinfo {pages} {012425} (\bibinfo {year} {2023})}\BibitemShut {NoStop}%
\bibitem [{\citenamefont {Ozawa}\ \emph {et~al.}(2019)\citenamefont {Ozawa}, \citenamefont {Price}, \citenamefont {Amo}, \citenamefont {Goldman}, \citenamefont {Hafezi}, \citenamefont {Lu}, \citenamefont {Rechtsman}, \citenamefont {Schuster}, \citenamefont {Simon}, \citenamefont {Zilberberg},\ and\ \citenamefont {Carusotto}}]{ozawa19}%
  \BibitemOpen
  \bibfield  {author} {\bibinfo {author} {\bibfnamefont {T.}~\bibnamefont {Ozawa}}, \bibinfo {author} {\bibfnamefont {H.~M.}\ \bibnamefont {Price}}, \bibinfo {author} {\bibfnamefont {A.}~\bibnamefont {Amo}}, \bibinfo {author} {\bibfnamefont {N.}~\bibnamefont {Goldman}}, \bibinfo {author} {\bibfnamefont {M.}~\bibnamefont {Hafezi}}, \bibinfo {author} {\bibfnamefont {L.}~\bibnamefont {Lu}}, \bibinfo {author} {\bibfnamefont {M.~C.}\ \bibnamefont {Rechtsman}}, \bibinfo {author} {\bibfnamefont {D.}~\bibnamefont {Schuster}}, \bibinfo {author} {\bibfnamefont {J.}~\bibnamefont {Simon}}, \bibinfo {author} {\bibfnamefont {O.}~\bibnamefont {Zilberberg}},\ and\ \bibinfo {author} {\bibfnamefont {I.}~\bibnamefont {Carusotto}},\ }\bibfield  {title} {\bibinfo {title} {Topological photonics},\ }\href {https://doi.org/10.1103/RevModPhys.91.015006} {\bibfield  {journal} {\bibinfo  {journal} {Rev. Mod. Phys.}\ }\textbf {\bibinfo {volume} {91}},\ \bibinfo {pages} {015006} (\bibinfo {year} {2019})}\BibitemShut {NoStop}%
\bibitem [{\citenamefont {Kang}\ \emph {et~al.}(2023)\citenamefont {Kang}, \citenamefont {Wei}, \citenamefont {Zhang},\ and\ \citenamefont {Dong}}]{kang23}%
  \BibitemOpen
  \bibfield  {author} {\bibinfo {author} {\bibfnamefont {J.}~\bibnamefont {Kang}}, \bibinfo {author} {\bibfnamefont {R.}~\bibnamefont {Wei}}, \bibinfo {author} {\bibfnamefont {Q.}~\bibnamefont {Zhang}},\ and\ \bibinfo {author} {\bibfnamefont {G.}~\bibnamefont {Dong}},\ }\bibfield  {title} {\bibinfo {title} {Topological photonic states in waveguide arrays},\ }\href {https://doi.org/https://doi.org/10.1002/apxr.202200053} {\bibfield  {journal} {\bibinfo  {journal} {Advanced Physics Research}\ }\textbf {\bibinfo {volume} {2}},\ \bibinfo {pages} {2200053} (\bibinfo {year} {2023})}\BibitemShut {NoStop}%
\bibitem [{\citenamefont {Su}\ \emph {et~al.}(1979)\citenamefont {Su}, \citenamefont {Schrieffer},\ and\ \citenamefont {Heeger}}]{su79}%
  \BibitemOpen
  \bibfield  {author} {\bibinfo {author} {\bibfnamefont {W.~P.}\ \bibnamefont {Su}}, \bibinfo {author} {\bibfnamefont {J.~R.}\ \bibnamefont {Schrieffer}},\ and\ \bibinfo {author} {\bibfnamefont {A.~J.}\ \bibnamefont {Heeger}},\ }\bibfield  {title} {\bibinfo {title} {Solitons in polyacetylene},\ }\href {https://doi.org/10.1103/PhysRevLett.42.1698} {\bibfield  {journal} {\bibinfo  {journal} {Phys. Rev. Lett.}\ }\textbf {\bibinfo {volume} {42}},\ \bibinfo {pages} {1698} (\bibinfo {year} {1979})}\BibitemShut {NoStop}%
\bibitem [{\citenamefont {Batra}\ and\ \citenamefont {Sheet}(2020)}]{batra20}%
  \BibitemOpen
  \bibfield  {author} {\bibinfo {author} {\bibfnamefont {N.}~\bibnamefont {Batra}}\ and\ \bibinfo {author} {\bibfnamefont {G.}~\bibnamefont {Sheet}},\ }\bibfield  {title} {\bibinfo {title} {Physics with coffee and doughnuts},\ }\href {https://doi.org/10.1007/s12045-020-0995-x} {\bibfield  {journal} {\bibinfo  {journal} {Resonance}\ }\textbf {\bibinfo {volume} {25}},\ \bibinfo {pages} {765} (\bibinfo {year} {2020})}\BibitemShut {NoStop}%
\bibitem [{\citenamefont {Almeida}(2018)}]{almeida18}%
  \BibitemOpen
  \bibfield  {author} {\bibinfo {author} {\bibfnamefont {G.~M.~A.}\ \bibnamefont {Almeida}},\ }\bibfield  {title} {\bibinfo {title} {Interplay between speed and fidelity in off-resonant quantum-state-transfer protocols},\ }\href {https://doi.org/10.1103/PhysRevA.98.012334} {\bibfield  {journal} {\bibinfo  {journal} {Phys. Rev. A}\ }\textbf {\bibinfo {volume} {98}},\ \bibinfo {pages} {012334} (\bibinfo {year} {2018})}\BibitemShut {NoStop}%
\bibitem [{\citenamefont {Ciccarello}(2011)}]{ciccarello11}%
  \BibitemOpen
  \bibfield  {author} {\bibinfo {author} {\bibfnamefont {F.}~\bibnamefont {Ciccarello}},\ }\bibfield  {title} {\bibinfo {title} {Resonant atom-field interaction in large-size coupled-cavity arrays},\ }\href@noop {} {\bibfield  {journal} {\bibinfo  {journal} {Phys. Rev. A}\ }\textbf {\bibinfo {volume} {83}},\ \bibinfo {pages} {043802} (\bibinfo {year} {2011})}\BibitemShut {NoStop}%
\bibitem [{Note1()}]{Note1}%
  \BibitemOpen
  \bibinfo {note} {It is a reasonable approximation given we are only interested in the statistics of the local wavefunction amplitude, rather than its exact time evolution.}\BibitemShut {Stop}%
\bibitem [{\citenamefont {Almeida}\ \emph {et~al.}(2016)\citenamefont {Almeida}, \citenamefont {Ciccarello}, \citenamefont {Apollaro},\ and\ \citenamefont {Souza}}]{almeida16}%
  \BibitemOpen
  \bibfield  {author} {\bibinfo {author} {\bibfnamefont {G.~M.~A.}\ \bibnamefont {Almeida}}, \bibinfo {author} {\bibfnamefont {F.}~\bibnamefont {Ciccarello}}, \bibinfo {author} {\bibfnamefont {T.~J.~G.}\ \bibnamefont {Apollaro}},\ and\ \bibinfo {author} {\bibfnamefont {A.~M.~C.}\ \bibnamefont {Souza}},\ }\bibfield  {title} {\bibinfo {title} {Quantum-state transfer in staggered coupled-cavity arrays},\ }\href {https://doi.org/10.1103/PhysRevA.93.032310} {\bibfield  {journal} {\bibinfo  {journal} {Phys. Rev. A}\ }\textbf {\bibinfo {volume} {93}},\ \bibinfo {pages} {032310} (\bibinfo {year} {2016})}\BibitemShut {NoStop}%
\bibitem [{\citenamefont {Campos~Venuti}\ \emph {et~al.}(2007)\citenamefont {Campos~Venuti}, \citenamefont {Giampaolo}, \citenamefont {Illuminati},\ and\ \citenamefont {Zanardi}}]{venuti07}%
  \BibitemOpen
  \bibfield  {author} {\bibinfo {author} {\bibfnamefont {L.}~\bibnamefont {Campos~Venuti}}, \bibinfo {author} {\bibfnamefont {S.~M.}\ \bibnamefont {Giampaolo}}, \bibinfo {author} {\bibfnamefont {F.}~\bibnamefont {Illuminati}},\ and\ \bibinfo {author} {\bibfnamefont {P.}~\bibnamefont {Zanardi}},\ }\bibfield  {title} {\bibinfo {title} {Long-distance entanglement and quantum teleportation in {$XX$} spin chains},\ }\href {https://doi.org/10.1103/PhysRevA.76.052328} {\bibfield  {journal} {\bibinfo  {journal} {Phys. Rev. A}\ }\textbf {\bibinfo {volume} {76}},\ \bibinfo {pages} {052328} (\bibinfo {year} {2007})}\BibitemShut {NoStop}%
\bibitem [{\citenamefont {Amico}\ \emph {et~al.}(2004)\citenamefont {Amico}, \citenamefont {Osterloh}, \citenamefont {Plastina}, \citenamefont {Fazio},\ and\ \citenamefont {Massimo~Palma}}]{amico04}%
  \BibitemOpen
  \bibfield  {author} {\bibinfo {author} {\bibfnamefont {L.}~\bibnamefont {Amico}}, \bibinfo {author} {\bibfnamefont {A.}~\bibnamefont {Osterloh}}, \bibinfo {author} {\bibfnamefont {F.}~\bibnamefont {Plastina}}, \bibinfo {author} {\bibfnamefont {R.}~\bibnamefont {Fazio}},\ and\ \bibinfo {author} {\bibfnamefont {G.}~\bibnamefont {Massimo~Palma}},\ }\bibfield  {title} {\bibinfo {title} {Dynamics of entanglement in one-dimensional spin systems},\ }\href {https://doi.org/10.1103/PhysRevA.69.022304} {\bibfield  {journal} {\bibinfo  {journal} {Phys. Rev. A}\ }\textbf {\bibinfo {volume} {69}},\ \bibinfo {pages} {022304} (\bibinfo {year} {2004})}\BibitemShut {NoStop}%
\bibitem [{\citenamefont {Haldane}(1983)}]{haldane83}%
  \BibitemOpen
  \bibfield  {author} {\bibinfo {author} {\bibfnamefont {F.~D.~M.}\ \bibnamefont {Haldane}},\ }\bibfield  {title} {\bibinfo {title} {Nonlinear field theory of large-spin heisenberg antiferromagnets: Semiclassically quantized solitons of the one-dimensional easy-axis n\'eel state},\ }\href {https://doi.org/10.1103/PhysRevLett.50.1153} {\bibfield  {journal} {\bibinfo  {journal} {Phys. Rev. Lett.}\ }\textbf {\bibinfo {volume} {50}},\ \bibinfo {pages} {1153} (\bibinfo {year} {1983})}\BibitemShut {NoStop}%
\bibitem [{\citenamefont {Oshikawa}\ \emph {et~al.}(1997)\citenamefont {Oshikawa}, \citenamefont {Yamanaka},\ and\ \citenamefont {Affleck}}]{oshikawa97}%
  \BibitemOpen
  \bibfield  {author} {\bibinfo {author} {\bibfnamefont {M.}~\bibnamefont {Oshikawa}}, \bibinfo {author} {\bibfnamefont {M.}~\bibnamefont {Yamanaka}},\ and\ \bibinfo {author} {\bibfnamefont {I.}~\bibnamefont {Affleck}},\ }\bibfield  {title} {\bibinfo {title} {Magnetization plateaus in spin chains: ``{H}aldane gap'' for half-integer spins},\ }\href {https://doi.org/10.1103/PhysRevLett.78.1984} {\bibfield  {journal} {\bibinfo  {journal} {Phys. Rev. Lett.}\ }\textbf {\bibinfo {volume} {78}},\ \bibinfo {pages} {1984} (\bibinfo {year} {1997})}\BibitemShut {NoStop}%
\bibitem [{\citenamefont {Ver\'{\i}ssimo}\ \emph {et~al.}(2021)\citenamefont {Ver\'{\i}ssimo}, \citenamefont {Pereira},\ and\ \citenamefont {Lyra}}]{verissimo21}%
  \BibitemOpen
  \bibfield  {author} {\bibinfo {author} {\bibfnamefont {L.~M.}\ \bibnamefont {Ver\'{\i}ssimo}}, \bibinfo {author} {\bibfnamefont {M.~S.~S.}\ \bibnamefont {Pereira}},\ and\ \bibinfo {author} {\bibfnamefont {M.~L.}\ \bibnamefont {Lyra}},\ }\bibfield  {title} {\bibinfo {title} {Tangential finite-size scaling at the gaussian topological transition in the quantum spin-1 anisotropic chain},\ }\href {https://doi.org/10.1103/PhysRevB.104.024409} {\bibfield  {journal} {\bibinfo  {journal} {Phys. Rev. B}\ }\textbf {\bibinfo {volume} {104}},\ \bibinfo {pages} {024409} (\bibinfo {year} {2021})}\BibitemShut {NoStop}%
\bibitem [{\citenamefont {Xu}\ \emph {et~al.}(2003)\citenamefont {Xu}, \citenamefont {Dai}, \citenamefont {Ying},\ and\ \citenamefont {Zheng}}]{xu03}%
  \BibitemOpen
  \bibfield  {author} {\bibinfo {author} {\bibfnamefont {Z.}~\bibnamefont {Xu}}, \bibinfo {author} {\bibfnamefont {J.}~\bibnamefont {Dai}}, \bibinfo {author} {\bibfnamefont {H.}~\bibnamefont {Ying}},\ and\ \bibinfo {author} {\bibfnamefont {B.}~\bibnamefont {Zheng}},\ }\bibfield  {title} {\bibinfo {title} {Successive valence-bond-state transitions in quantum mixed spin chains},\ }\href {https://doi.org/10.1103/PhysRevB.67.214426} {\bibfield  {journal} {\bibinfo  {journal} {Phys. Rev. B}\ }\textbf {\bibinfo {volume} {67}},\ \bibinfo {pages} {214426} (\bibinfo {year} {2003})}\BibitemShut {NoStop}%
\bibitem [{\citenamefont {Nevado}\ \emph {et~al.}(2017)\citenamefont {Nevado}, \citenamefont {Fern\'andez-Lorenzo},\ and\ \citenamefont {Porras}}]{nevado17}%
  \BibitemOpen
  \bibfield  {author} {\bibinfo {author} {\bibfnamefont {P.}~\bibnamefont {Nevado}}, \bibinfo {author} {\bibfnamefont {S.}~\bibnamefont {Fern\'andez-Lorenzo}},\ and\ \bibinfo {author} {\bibfnamefont {D.}~\bibnamefont {Porras}},\ }\bibfield  {title} {\bibinfo {title} {Topological edge states in periodically driven trapped-ion chains},\ }\href {https://doi.org/10.1103/PhysRevLett.119.210401} {\bibfield  {journal} {\bibinfo  {journal} {Phys. Rev. Lett.}\ }\textbf {\bibinfo {volume} {119}},\ \bibinfo {pages} {210401} (\bibinfo {year} {2017})}\BibitemShut {NoStop}%
\bibitem [{\citenamefont {Youssefi}\ \emph {et~al.}(2022)\citenamefont {Youssefi}, \citenamefont {Kono}, \citenamefont {Bancora}, \citenamefont {Chegnizadeh}, \citenamefont {Pan}, \citenamefont {Vovk},\ and\ \citenamefont {Kippenberg}}]{youssefi22}%
  \BibitemOpen
  \bibfield  {author} {\bibinfo {author} {\bibfnamefont {A.}~\bibnamefont {Youssefi}}, \bibinfo {author} {\bibfnamefont {S.}~\bibnamefont {Kono}}, \bibinfo {author} {\bibfnamefont {A.}~\bibnamefont {Bancora}}, \bibinfo {author} {\bibfnamefont {M.}~\bibnamefont {Chegnizadeh}}, \bibinfo {author} {\bibfnamefont {J.}~\bibnamefont {Pan}}, \bibinfo {author} {\bibfnamefont {T.}~\bibnamefont {Vovk}},\ and\ \bibinfo {author} {\bibfnamefont {T.~J.}\ \bibnamefont {Kippenberg}},\ }\bibfield  {title} {\bibinfo {title} {Topological lattices realized in superconducting circuit optomechanics},\ }\href {https://doi.org/10.1038/s41586-022-05367-9} {\bibfield  {journal} {\bibinfo  {journal} {Nature}\ }\textbf {\bibinfo {volume} {612}},\ \bibinfo {pages} {666} (\bibinfo {year} {2022})}\BibitemShut {NoStop}%
\bibitem [{\citenamefont {Zurita}\ \emph {et~al.}(2023)\citenamefont {Zurita}, \citenamefont {Creffield},\ and\ \citenamefont {Platero}}]{zurita23}%
  \BibitemOpen
  \bibfield  {author} {\bibinfo {author} {\bibfnamefont {J.}~\bibnamefont {Zurita}}, \bibinfo {author} {\bibfnamefont {C.~E.}\ \bibnamefont {Creffield}},\ and\ \bibinfo {author} {\bibfnamefont {G.}~\bibnamefont {Platero}},\ }\bibfield  {title} {\bibinfo {title} {Fast quantum transfer mediated by topological domain walls},\ }\href {https://doi.org/10.22331/q-2023-06-22-1043} {\bibfield  {journal} {\bibinfo  {journal} {{Quantum}}\ }\textbf {\bibinfo {volume} {7}},\ \bibinfo {pages} {1043} (\bibinfo {year} {2023})}\BibitemShut {NoStop}%
\end{thebibliography}

%

\end{document}